\documentclass[11pt]{article}
\usepackage{amssymb}
\usepackage{amsmath}
\usepackage{amsfonts}
\newcommand{\Z}{\mathbb{Z}}

\newcommand{\be}{\begin{equation}}
\newcommand{\ee}{\end{equation}}
\newcommand{\bea}{\begin{eqnarray}}
\newcommand{\eea}{\end{eqnarray}}
\newcommand{\nn}{\nonumber}
\newcommand{\kt}{\rangle}

\newcommand{\ed}{\end{document}}

\newcommand{\Q}{{\cal Q}}
\begin{document}
\title{A New Interpretation for Orthofermions}
\author{Keivan Aghababaei Samani\thanks{E-mail address: samani@cc.iut.ac.ir}
\\ \\
{\it Department of Physics, Isfahan University of Technology (IUT)},\\
{\it Isfahan 84154, Iran}}
\date{ }
\maketitle
\begin{abstract}
In this article we introduce a simple physical model which realizes the algebra of
orthofermions. The model is constructed from a cylinder which can be filled with some
balls. The creation and annihilation operators of orthofermions are related to the
creation and annihilation operators of  balls in  certain positions in the cylinder.
Relationship between this model and topological symmetries in quantum
mechanics\cite{p1} is investigated.
\end{abstract}
Keywords: Orthofermions, Topological Symmetries.
\newpage
\section*{Introduction}
The study of general types of statistics dates back to 1940's. Since then there have
been many attempts to generalize the concept of bosons and fermions. Some of the most
well known generalized statistics are of anyons \cite{anyon1,anyon}, parafermions and
parabosons \cite{green,para}, and orthofermions and orthobosons \cite{ortho}. These
generalized types of statistics are discussed in different contexts. For examples
parafermions and parabosons was originally introduced as general types of fields
(parafields) \cite{green,para}. Then they found some applications in generalization
of supersymmetric quantum mechanics namely parasupersymmetric quantum mechanics
\cite{rs,bd}. Orthofermi and orthobose statistics originally introduced as new types
of statistics \cite{ortho} and subsequently used for another generalization of
supersymmetric quantum mechanics calld orthosupersymmetric quantum mechanics
\cite{orthosusy}. They also turned out to be the origin of some topological
symmetries \cite{p3} which we have already introduced and explored their algebra
\cite{p1,p2}. A topological symmetry is a generalization of supersymmetry from a
topological point of view, namely the Witten index\cite{witten82}.

Orthofermi statistics is a generalization of fermi statistics in the sense that an
orbital state can not contain more than one particle regardless of its spin.

As is well known, particles with statistics other than bose or fermi statistics  have
not been seen yet. But the concept of generalized statistics may be still useful as
we can construct some  models which obey their algebras. The goal of this paper is to
introduce a model, called the cylinder model, which realizes the algebra of
orthofermions.

The statistics of orthofermions of order $p$ is given by the following equations
\cite{ortho}
   \bea
   &&c_\alpha c_\beta^\dagger +\delta_{\alpha \beta} \sum^p_{\gamma=1}
   c_\gamma^\dagger c_\gamma=\delta_{\alpha \beta}\;,
   \label{of1}\\
   &&c_\alpha c_\beta=0\;.
   \label{of2}
   \eea
where $c_\alpha$ and $c_\alpha^\dagger$ are annihilation and creation operators
respectively. The above algebra is a generalization of fermions in the sense that for
$p=1$ we get the fermionic algebra
   \bea
   &&c_1 c_1^\dagger +c_1^\dagger c_1=1\;,
   \label{f1}\\
   &&c_1^2=0\;.
   \label{f2}
   \eea
The representation of $c_\alpha$ (up to a unitary equivalence)  is given by the
$(p+1)\times (p+1)$ matrices
   \be
   [c_\alpha]_{ij}=\delta_{i,1}\delta_{j,\alpha+1}\;,~~ i,j=1,\cdots,p+1\;.
   \label{4.3}
   \ee
\section*{The Cylinder Model}
Consider a cylinder of radius $r$ and length $pr$, where $p$ is a non negative
integer. Suppose that  this cylinder is put vertically on a surface so that one can
drop some balls of radius $r$ in it. Clearly one  can drop at most $p$ balls in the
cylinder. To empty the cylinder one should take the balls out of the cylinder in the
opposite manner it is filled, i.e. the last ball put in the cylinder should be the
first which will be taken out of it. (This cylinder is very similar to some type of
memories in computer called {\bf LIFO} stack. {\bf LIFO} means that the {\bf L}ast
object we put {\bf I}n the stack is the {\bf F}irst one we take {\bf O}ut of it).

Let's label the positions of balls in the cylinder from the bottom of the cylinder
with numbers $1,2,\cdots , p$. We denote the state of the cylinder containing
$\alpha$ balls, $\alpha=0,1,\cdots p$, by $|\alpha\kt$. Therefore $|0\kt$ is the
state of empty cylinder (the vacuum state).

Furthermore let $b_\alpha^\dagger$ and $b_\alpha$ be the creation and annihilation
operators of a ball in position $\alpha (\alpha \in \{ 1,2,\cdots, p\})$
respectively.  So we define
    \be
    |\alpha\kt=b_\alpha^\dagger
    b_{\alpha-1}^\dagger\cdots b_1^\dagger |0\kt\;,
    \label{2}
    \ee

The vacuum state $|0\kt$ should, by its definition, satisfy the following equations
    \bea
    &&b_\alpha|0\kt=0~~,~~\alpha=1,2,\cdots,p
    \label{3.1}\\
    &&b_\alpha^\dagger|0\kt=0~~,~~\alpha=2,3,\cdots,p
    \label{3.2}\\
    &&b_1^\dagger|0\kt=|1\kt
    \label{3.3}
    \eea

Using operators $b_\alpha^\dagger$ and $b_\alpha$ one can drop a ball in the cylinder
or draw  a ball out of it respectively considering the following rules:
\begin{itemize}
\item[1.] A ball can be created in position $\alpha$ if and only
if the position $\alpha$ is already empty and all positions $1,2,\cdots,\alpha-1$ are
filled with balls.
\item[2.] A ball can be annihilated in position $\alpha$ if and
only if all positions $\alpha+1,\cdots, p$ are empty.
\end{itemize}

The above two rules is summarized  in the following algebraic relations between
creation and annihilation operators
    \bea
    &&b_\beta^\dagger b_\alpha^\dagger=b_\alpha b_\beta=0\;,~~ \beta \ne\alpha+1
    \label{balgebra1}\\
    &&b_\beta^\dagger b_\alpha=b_\alpha b_\beta^\dagger=0\;,~~ \beta\ne\alpha
    \label{balgebra2}
    \eea

Now let see what is the meaning of $b_\alpha^\dagger b_\alpha$ and $b_\alpha
b_\alpha^\dagger$. In view of Eq.~(\ref{2}), (\ref{balgebra1}), (\ref{balgebra2}) it
is easy to see that
    \bea
    && b_\alpha^\dagger b_\alpha|\beta\kt=0\;,~~\beta\ne\alpha\;,
    \label{2.1}\\
    && b_\alpha
    b_\alpha^\dagger|\beta\kt=0\;,~~\beta\ne\alpha-1\;.
    \label{2.2}
    \eea
So $b_\alpha^\dagger b_\alpha$ and $b_\alpha b_\alpha^\dagger$ are projection
operators on subspaces spanned by $|\alpha\kt$ and $|\alpha-1\kt$ respectively.
Because the values of $\alpha$ are constrained to the set $\{1,2,\cdots,p\}$,
$b_\alpha^\dagger b_\alpha$ can not project on the subspace spanned by $|0\kt$. The
same thing is true for $b_\alpha b_\alpha^\dagger$ and $|p\kt$. Consequently
    \be
    \sum^p_{\alpha=1}b_\alpha^\dagger b_\alpha +b_1 b_1^\dagger=\sum^p_{\alpha=1}
    b_\alpha b_\alpha^\dagger +b_p^\dagger b_p=1\;,
    \label{bg3}
    \ee

One can also verify that the operator $N:=\sum^p_{\alpha=1}b_\alpha^\dagger b_\alpha$
shows that the cylinder is empty or not (if its eigenvalue on a state is zero that
state correspond to empty cylinder; otherwise the cylinder is not empty). In the same
manner $\sum^p_{\alpha=1}b_\alpha b_\alpha^\dagger$ shows that the cylinder is
completely  full or not.

Next we derive some more algebraic relations between $b_\alpha$s and
$b_\alpha^\dagger$s which will be used in next section to construct the algebra of
orthofermions. Eq.~(\ref{bg3}) together with Eqs.~(\ref{balgebra1}) and
(\ref{balgebra2}) results in
    \bea
    && b_\beta = b_\beta b_\beta^\dagger b_\beta=b_{\beta-1}^\dagger b_{\beta-1} b_\beta
    =b_\beta b_{\beta+1} b_{\beta+1}^\dagger\;,
    \label{bg4}\\
    && b_\beta^\dagger = b_\beta^\dagger b_\beta b_\beta^\dagger=b_\beta^\dagger
    b_{\beta-1}^\dagger b_{\beta-1}=b_{\beta+1}b_{\beta+1}^\dagger b_\beta^\dagger\;.
    \label{bg5}
    \eea
\section*{Orthofermions and the Cylinder
Model}
Let us define a new set of creation and annihilation operators,
$c_\alpha^\dagger$ and $c_\alpha$
    \bea
    &&c_\alpha^\dagger:=b_\alpha^\dagger \cdots b_1^\dagger\;,
    \label{4.1}\\
    &&c_\alpha:=b_1\cdots b_\alpha\;,
    \label{4.2}
    \eea
$c_\alpha^\dagger$ fills the cylinder with $\alpha$ balls starting position $1$. In
view of Eq.~(\ref{balgebra1}) and definitions~(\ref{4.1}) and (\ref{4.2}) one can
easily see that
    \be
    c_\alpha^\dagger c_\beta^\dagger=c_\alpha c_\beta=0\;,~ \alpha,\beta= 1,2,\cdots, p\;.
    \label{5}
    \ee
This equation is the first equation of orthofermions algebra, namely Eq.~(\ref{of1}).
The Physical meaning of this equation is that the cylinder can not contain $\alpha$
and $\beta$ balls in the same time and if we fill it with $\alpha$ balls, we can not
put another $\alpha$ balls in the same positions in the cylinder again.

One can also use Eqs.~(\ref{bg4})-(\ref{4.2})  to get
    \bea
    &&c_\alpha^\dagger c_\alpha=b_\alpha^\dagger
    b_\alpha~~,\alpha=1,\cdots,p
    \label{5.1}\\
    &&c_\alpha c_\alpha^\dagger =b_1b_1^\dagger
    ~~,\alpha=1,\cdots,p
    \label{5.2}\\
    &&c_\alpha c_\beta^\dagger =0~~,\beta\ne \alpha
    \label{5.3}
    \eea
Putting the above equations together one arrives at the first equation of
orthofermions algebra namely Eq.~(\ref{of1}).

Before leaving this section, let us have a look at the inverse of Eqs~(\ref{4.1}) and
(\ref{4.2}). Using Eqs.~(\ref{bg4}) and (\ref{bg5}), it can be easily verified that
the inverse relations are
    \bea
    &&b_\alpha^\dagger=c_\alpha^\dagger  c_{\alpha-1}\;,
    \label{4.1b}\\
    &&b_\alpha=c_{\alpha-1}^\dagger c_\alpha\;,
    \label{4.2b}
    \eea
for  $\alpha>1$, and $b_1=c_1$. Thus the relations between $b_\alpha$'s and
$c_\alpha$'s are invertible and therefore having the representation of $c_\alpha$'s
in hand, the representation of $b_\alpha$'s is determined  uniquely and vice versa.
\section*{Relation to Topological Symmetries}
As it is mentioned in Introduction topological symmetries are generalizations of
supersymmetry. Their physical properties and algebra are investigated in Refs.
\cite{p1,p2}. Here we are going to have a look at the relation between cylinder model
and topological symmetries.

The algebra of a $\Z_2$-graded topological symmetry of type $(1,p)$ is given by the
following equations
    \bea
    &&[H,\Q]=0\;,
    \label{ts1}\\
    &&\{\Q^2,\Q^\dagger\}+\Q\Q^\dagger\Q=2H\Q\;,
    \label{ts2}\\
    &&\Q^3=0\;,
    \label{ts3}\\
   &&[H,\tau]=\{\tau,\Q\}=0\;,
   \label{ts4}\\
   && \tau^2=1\;,~~ \tau^\dagger=\tau\;,
    \label{t2}
    \eea
where $H$ is the Hamiltonian of the system, $\Q$ is the symmetry generator, and
$\tau$ is the grading operator.

Now Consider a quantum system with the Hamiltonian
   \be
   H=a^\dagger a+\sum^p_{\gamma=1}c_\gamma^\dagger c_\gamma\;,
   \label{4.11}
   \ee
where $a$ is the annihilation operator for a bosonic degree of freedom satisfying
$[a,a^\dagger ]=1\;,$ and $c_\gamma\;$, with $\gamma=1,\cdots,p$, are annihilation
operators of orthofermions of order $p$. One can easily verify that this Hamiltonian
together with symmetry generator\footnote{It should be mentioned that one  can
equivalently take the more general symmetry generator $\Q=\frac{1}{\sqrt {2p}}\left(a
\sum^r_{j=1}c^\dagger_{\gamma_j} +a^\dagger \sum^p_{j=r+1} c_{\gamma_j}\right)\;,$
where $(\gamma_1,\cdots\gamma_p)$ is an arbitrary permutation of $(1,\cdots,p)$ and
$r$ is an integer between $1$ and $p-1$.}
   \be
   \Q=\frac{1}{\sqrt {2p}}\left(a \sum^r_{j=1}c^\dagger_j +a^\dagger \sum^p_{j=r+1}
   c_j\right)\;,
   \label{4.6}
   \ee
satisfy  the algebra of a topological $\Z_2$-graded symmetry of type $(1,p)$,
provided that the grading operator is given by $\tau=(-1)^N$ where
$N:=\sum^p_{\alpha=1}c_\alpha^\dagger c_\alpha$\cite{p3}. Now coming back to the
cylinder model one can see that the operator $N$ indicates that the cylinder is empty
or not. comparing $\tau$ with operator $(-1)^F$ in supersymmetry one can see that $N$
is a generalization of fermionic number operator $F$, and this has a clear meaning in
the cylinder model of orthofermions.

The algebra of ($\Z_n$-graded) topological symmetry of type
$(\underbrace{1,1,\cdots,1}_{n~{\rm times}})$ is given by
    \bea
   &&\Q^n=K\;, \label{2.3}\\
    &&Q_1^n +M_{n-2} Q_1^{n-2} +\cdots
    =({\frac{1}{\sqrt 2}})^n (K+K^\dagger)\;,
    \label{2.4}\\
    &&Q_2^n +M_{n-2} Q_2^{n-2} +\cdots
    =({\frac{1}{\sqrt 2}})^n (i^n K^\dagger+ (-i)^n K) \;,
    \label{2.5}\\
   &&[\tau,\Q]_q=0\;.
   \label{2.6}
    \eea
where
   \be
    Q_1:=\frac{1}{\sqrt{2}}\:(\Q+\Q^\dagger)\;~~~
    {\rm and}~~~Q_2:=\frac{-i}{\sqrt{2}}\:(\Q-\Q^\dagger)\;.
    \nn
    \ee
$M_i$s and $K$ are operators commuting with all other operators, $M_i$ and $K$ are
Hermitian and $\tau$ is the grading operator satisfying
    \bea
    \tau^n&=&1\;,
    \label{tn1}\\
    \tau^\dagger&=&\tau^{-1}\;,
    \label{tn2}\\
    \left[H,\tau\right]&=&0\;.
    \label{tn3}
    \eea
Here $[.,.]_q$ stands for {\it $q$-commutator} defined by
$[O_1,O_2]_q:=O_1O_2-qO_2O_1$, and $q:=e^{2\pi i/n}$. It is easily verified that
Hamiltonian (\ref{4.11}) together with symmetry generator
  \be
   \Q=ac_1^\dagger +c_2^\dagger c_1+\cdots+c_p^\dagger c_{p-1}+a^\dagger c_p\;.
   \label{4.8}
   \ee
satisfy the above algebra for $n=p+1$~\cite{p3}.  The grading operator is
   \be
   \tau=q^{\cal N},~~~{\rm where}~~~{\cal N}:=\sum^p_{\alpha=1}\alpha N_\alpha\;,
   \label{4.9}
   \ee
and the operator $K$ of Eq.~(\ref{2.3}) is identified with the Hamiltonian $H$. The
operators $M_i$ of Eqs.~(\ref{2.4}) and (\ref{2.5}) are given in terms of $H$
according to
        \be
        M_{n-2k}=(-1)^k\left[\frac{1}{2^k} {{n-k-1} \choose k}+
        \frac{1}{2^{k-1}}{{n-k-1} \choose {k-1}}H\right]\;,
        \label{4.10}
        \ee
and $\left(\begin{array}{c}a\\b\end{array}\right):=\frac{a!}{b!(a-b)!}$.

In our cylinder model the (number) operator ${\cal N}:=\sum^p_{\alpha=1}\alpha
c_\alpha^\dagger c_\alpha$ has a nice meaning. It counts the number of balls in the
cylinder. This is another generalization of fermionic number operator.

\section*{Conclusion}
In this article we introduced a simple physical model which realizes the algebra of
orthofermions. We explored how one can get the orthofermionic algebra using the
creation and annihilation operators of some balls in certain positions of a cylinder.
We also addressed the relationship between two number operators
$N=\sum^p_{\alpha=1}c_\alpha^\dagger c_\alpha$ and ${\cal N}:=\sum^p_{\alpha=1}\alpha
c_\alpha^\dagger c_\alpha$ and grading operators of two kinds of topological
symmetries namely $\Z_2$-graded topological symmetry of type $(1,p)$ and
$\Z_{p+1}$-graded topological symmetry of type $(1,1,\cdots,1)$.

These relationships may shed some light on the meaning of topological symmetries as
generalizations of supersymmetry. They also show how orthofermions  may be the basic
ingredients of topological symmetries.

\section*{Acknowledgment}
I wish to thank F. Loran for fruitful discussions and comments. Financial supports of
Isfahan University of Technology is acknowledged.

\end{document}